\documentclass[journal]{IEEEtran}
\usepackage{amsmath}
\usepackage{bm}
\usepackage{amssymb}
\usepackage{graphicx}
\usepackage{color,colortbl} 
\usepackage[dvipsnames]{xcolor}
\usepackage{algpseudocode,algorithm}
\usepackage{hyperref}
\usepackage{enumerate}
\usepackage{boldline}
\usepackage{multicol,lipsum}
\usepackage[shortlabels]{enumitem}

\usepackage[caption=false,font=footnotesize]{subfig}
\newcommand{\splitatcommas}[1]{%
  \begingroup
  \ifnum\mathcode`,="8000
  \else
    \begingroup\lccode`~=`, \lowercase{\endgroup
      \edef~{\mathchar\the\mathcode`, \penalty0 \noexpand\hspace{0pt plus 1em}}%
    }\mathcode`,="8000
  \fi
  #1%
  \endgroup
}

\usepackage{cite}
\usepackage{url}
\begin{document}

\title{Cooperative mmWave PHD-SLAM with Moving Scatterers}
\author{
\IEEEauthorblockN{Hyowon Kim, \IEEEmembership{Member, IEEE}, Jaebok Lee, \IEEEmembership{Student Member, IEEE}, Yu Ge, \IEEEmembership{Student Member, IEEE}, Fan Jiang, \IEEEmembership{Member, IEEE}, Sunwoo Kim, \IEEEmembership{Senior Member, IEEE}, and Henk Wymeersch, \IEEEmembership{Senior Member, IEEE}}
\thanks{H.~Kim, J.~Lee, and S.~Kim are with the Department of Electronic Engineering, Hanyang University, 04763 Seoul, Republic of Korea (email: khw870511@hanyang.ac.kr; ok7393@hanyang.ac.kr; remero@hanyang.ac.kr).}
\thanks{Y.~Ge, F.~Jiang, and H.~Wymeersch are with the Department of Electrical Engineering, Chalmers University of Technology, 412 58 Gothenburg, Sweden (email: yuge@chalmers.se; fan.jiang@chalmers.se; henkw@chalmers.se).}
\thanks{This work was supported in part by the MSIP, Korea, under the ITRC support program (IITP-2020-2017-0-01637) supervised by the IITP, and by the Wallenberg AI, Autonomous Systems and Software Program (WASP) funded by the Knut and Alice Wallenberg Foundation, the Vinnova 5GPOS project under grant 2019-03085, and the European Commission through the H2020 project Hexa-X (Grant Agreement no. 101015956).}
}

\maketitle

\begin{abstract}
    Using the multiple-model~(MM) probability hypothesis density~(PHD) filter, millimeter wave~(mmWave) radio simultaneous localization and mapping~(SLAM) in vehicular scenarios is susceptible to movements of objects, in particular vehicles driving in parallel with the ego vehicle. 
    We propose and evaluate two countermeasures to track vehicle scatterers~(VSs) in mmWave radio MM-PHD-SLAM.
    First, locally at each vehicle, we generate and treat the VS map PHD in the context of Bayesian recursion, and modify vehicle state correction with the VS map PHD.
    Second, in the global map fusion process at the base station, we average the VS map PHD and upload it with self-vehicle posterior density, compute fusion weights, and prune the target with low Gaussian weight in the context of arithmetic average-based map fusion.
    From simulation results, the proposed cooperative mmWave radio MM-PHD-SLAM filter is shown to outperform the previous filter in VS scenarios.

\end{abstract}

\section{Introduction}
    In millimeter wave~(mmWave) signals, it is possible to obtain highly resolvable channel parameters in time and angular domains~\cite{Henk_5GmmWavePosi_WC2018}, enabling accurate mapping of landmarks.
    Therefore, a variety of radio simultaneous localization and mapping~(SLAM) works have been recently developed~\cite{Erik_BPSLAM_TWC2019,Rico_BPSLAM_JSTSP2019,Hyowon_TWC2020}. A single type of static landmark was considered in~\cite{Erik_BPSLAM_TWC2019,Rico_BPSLAM_JSTSP2019}.
    However, since radio environments contain different types of objects (small scattering objects, large reflecting objects, and moving objects), with different state definitions, multiple model~(MM) filters, such as the probability hypothesis density~(PHD) filter have been developed~\cite{Pasha_MMPHD_TAES2009,Karl_MMPHD_FUSION2014} in the context of random finite set~(RFS) theory~\cite{mahler_book_2007}.
    Such MM-PHD-SLAM was applied to mmWave radio SLAM in~\cite{Hyowon_TWC2020}, treating transient targets as clutter in~\cite{Hyowon_TWC2020}, under the assumption they are only visible for short intervals.

\begin{figure}
\begin{centering}
	\includegraphics[width=.95\columnwidth]{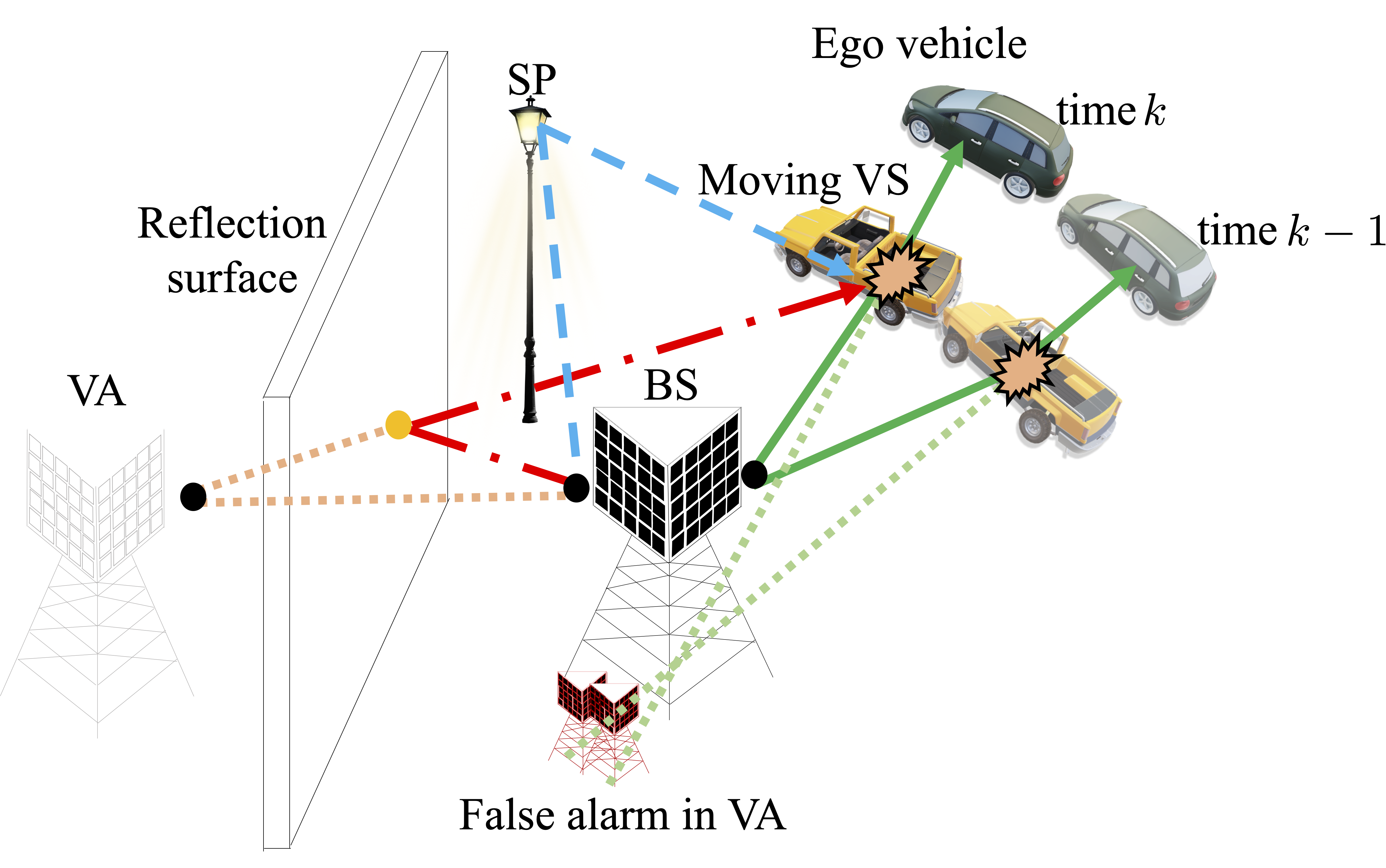}
	\caption{Channel propagation in mmWave signals, depending on static landmarks~(e.g., virtual anchors (VAs) and scatter points (SPs)) and moving vehicle scatterers (VSs). When vehicles are driving in parallel, false alarm occurs in the virtual anchor map.}
	\label{Fig:VS_channel}
\end{centering}
\end{figure}

    %
    In vehicular mmWave networks, the problem of moving objects is particularly relevant: a vehicle mmWave receiver can detect scattered signals by neighboring vehicles with the dynamics for long periods of time \cite{bas2017dynamic}, called as vehicle scatterer~(VS) in this work.
    Therefore, these VSs cannot be modeled as clutter anymore, leading to incorrect mapping results, even under global map fusion at the base station (BS).
    For example, in  Fig.~\ref{Fig:VS_channel}, a VS is detected as a virtual anchor~(VA)\footnote{The VA indicates a mirror image, reflected by large reflecting objects, of the BS~\cite{Hyowon_TWC2020}.} due to the fundamental limitation in VS estimation, since the VS velocity cannot be obtained from the mmWave measurements.
    
    In this paper, we infer both moving VSs and static landmarks in mmWave radio MM-PHD-SLAM. 
    To handle the aforementioned challenge, we develop an extension of mmWave radio MM-PHD-SLAM~\cite{Hyowon_TWC2020} that accounts for moving targets and can track them under most conditions, except for fundamentally unidentifiable conditions (e.g., when vehicles are driving in parallel). Our main contributions are (i) a local countermeasure in the MM-PHD-SLAM filter at each vehicle, treating the VS map PHD in the context of Bayesian recursion, and modifying the vehicle state correction with the VS map PHD; (ii) a global countermeasure during global map fusion process at the BS, modifying asynchronous map fusion for the VS PHD, and computing both fusion weights and fused map PHDs. 
    

\section{System Model}\label{sec:SystemModel}

    We consider a mmWave propagation channel wherein static landmarks (a single BS, several VAs, and several scatter points (SPs), indexed by $i$) and moving vehicles are present. Moving vehicles can be mmWave receivers (indexed by $n$) or VSs (indexed by $i$). Note that a vehicle $n$ will be a VS indexed by $i$ for another mmWave receiver. 
    The BS periodically transmits a mmWave signal, reflected by specular surfaces~(specifying VAs), scattered by SPs and VSs before reaching the receiver on each mmWave-equipped vehicle. 
    
\subsubsection{State Definitions}
    We denote the vehicle state by ${\mathbf{s}_k^{n}=[(\mathbf{x}_k^{n})^\top, \alpha_k^{n}, \zeta_k^{n}, \xi_k^{n} ,B_k^{n}]^\top} \in \mathbb{R}^{7}$, where $\mathbf{x}_k^{n}=[x_{k}^{n},y_{k}^{n},z_{k}^{n}]^\top$, $\alpha_k^{n}$, $\zeta_k^{n}$, $\xi_k^{n}$, and $B_k^{n}$ are respectively the location, heading, translation speed, turn-rate, and clock bias.
    The targets (static landmarks and moving VSs) are
    modeled as an RFS
    $\mathcal{X}_k=\{(\mathbf{t}_k^1,m^1),...,(\mathbf{t}_k^{I_k},m^{I_k})\}$ with  set density $f(\mathcal{X}_k)$~\cite{mahler_book_2007}.
    The target index is denoted by $i$, and the state of target $i$ is denoted by $\mathbf{t}_k^i=[(\mathbf{x}_k^{i})^\top, (\mathbf{v}_k^{i})^\top, \xi_k^{i}]^\top \in \mathbb{R}^{7}$, where $\mathbf{v}_k^{i}\in \mathbb{R}^3$
    is the velocity. The type of target $i$ is denoted by $m^i \in \mathcal{M}= \{\text{BS},\text{VA},\text{SP},\text{VS}\}$, and $I_k=\lvert \mathcal{X}_k \rvert$, where $\lvert \cdot \rvert$ is the set cardinality.
    If $m\in\{\text{BS},\text{VA},\text{SP}\}$, $\mathbf{v}_k^i = \bm{0}^\top$ and $\xi_k^i = 0$.
    
    \subsubsection{Dynamical Models}
    
    With the known transition density $f_{\text{V}}(\mathbf{s}_{k}^{n}|\mathbf{s}_{k-1}^{n})$, the vehicle dynamics are modeled as
    \begin{align}\label{eq:VehicleDynamic}
        \mathbf{s}_k^{n} = \mathsf{m}_{\text{V}}(\mathbf{s}_{k-1}^{n}) + \mathbf{q}_k,
    \end{align}
    where $\mathsf{m}_{\text{V}}(\cdot)$ is the known transition function, and $\mathbf{q}_k\sim \mathcal{N}(\bm{0},\mathbf{Q})$, with 
    known covariance matrix $\mathbf{Q}$.
    For VS $i$, 
    with  known transition density $f_{\text{VS}}(\mathbf{t}_k^i|\mathbf{t}_{k-1}^i)$, 
    the single-target dynamics are modeled as
\begin{align}\label{eq:TragetDynamic}
    \mathbf{t}_k^i =
    \mathsf{m}_\text{VS}
    (\mathbf{t}_{k-1}^i) + \bar{\mathbf{q}}_k,
\end{align}
    where $\mathsf{m}_\text{VS}(\cdot)$ denotes the known transition function and $\bar{\mathbf{q}}_k\sim\mathcal{N}(\bm{0},\bar{\mathbf{Q}})$ with known covariance $\bar{\mathbf{Q}}$. Note that for SPs and VAs, $f_{\text{VS}}(\mathbf{t}_k^i|\mathbf{t}_{k-1}^i)=\delta(\mathbf{t}_k^i-\mathbf{t}_{k-1}^i)$. In general, $f_{\text{VS}}(\mathbf{t}_k^i|\mathbf{t}_{k-1}^i)$ and $f_{\text{V}}(\mathbf{s}_{k}^{n}|\mathbf{s}_{k-1}^{n})$ may differ, as the former represents the assumed model of VSs, which may be simplified compared to the real dynamics $f_{\text{V}}(\mathbf{s}_{k}^{n}|\mathbf{s}_{k-1}^{n})$.

\subsubsection{Measurement Models}


    
    Each receiver at vehicles can detect signals coming from targets $(\mathbf{t}_k^i,m^i) \in \mathcal{X}$.
    Signal detection depends on a certain detection probability, denoted by $\mathsf{p}_{\text{D},k}(\mathbf{s}_k^n,\mathbf{t}_k^i,m^i) \in [0,1]$, within the field-of-view~(FOV)~\cite{Henk_FOV_6G2020}.
    Using the channel estimation routine \cite{Yu_DiffusePMBM_Sensors2020}, vehicle $n$ obtains an un-ordered measurement set $\mathcal{Z}_k^n=\{\mathbf{z}_k^{n,1},...,\mathbf{z}_k^{n,J}\}$, and $J = \lvert \mathcal{Z}_k^n \rvert$.
    Either false alarms (e.g., channel estimation error) or short time visible transient targets (e.g., people) are regarded as clutter.
    Clutter is modeled as random, where the number of clutter measurements follows the Poisson distribution.
    Note that there is no identifier which measurement originates from which target or clutter.
    The measurement index is denoted by $j$. Following \cite{Hyowon_TWC2020,Yu_DiffusePMBM_Sensors2020} non-clutter measurements $\mathbf{z}_k^j$ are modeled as
\begin{align} \label{eq:measModel}
    \mathbf{z}_{k}^j = \mathsf{h}(\mathbf{s}_k, \mathbf{x}_k, m) + \mathbf{r}_{k}^j,
\end{align}
    where $\mathsf{h}(\mathbf{s}_k, \mathbf{x}_k, m) = [\tau_{k}^j, (\bm{\theta}_{k}^j)^\top, (\bm{\phi}_{k}^j)^\top]^\top$, and $\mathbf{r}_{k}^j\sim \mathcal{N}(\bm{0},\mathbf{R})$ denotes the measurement noise.
    Here, $\tau_{k}^j$, $\bm{\theta}_{k}^j$, $\bm{\phi}_{k}^j$, and $\mathbf{R}$ respectively denote a time-of-arrival (TOA), angle-of-arrival (AOA) in azimuth and elevation, angle-of-departure (AOD) in azimuth and elevation, and the known measurement covariance matrix.
    Channel parameters for BS, VAs, and SPs follow the geometric relations in~\cite[Appendix B]{Hyowon_TWC2020}, and VSs are also scatterer and thus the relation of VS is the same as that of SP. Note that from \eqref{eq:measModel}, the velocity and turn-rate cannot be determined. 
    

\section{Recap of Cooperative MM-PHD-SLAM Filter} \label{sec:PreviousPHDSLAM}
    The MM-PHD-SLAM filter from~\cite{Hyowon_TWC2020} consisting of local filter and global map fusion is briefly described, with $\overline{\mathcal{M}}= \{\text{BS},\text{VA},\text{SP}\}$. 
    Additional details can be found in~\cite[Sec.~IV,~V]{Hyowon_TWC2020}.
    
\subsection{Local PHD-SLAM}\label{sec:LocalPHD}
    For a single vehicle, PHD-SLAM is implemented by the Rao-blackwellized particle filter~(RBPF), and the vehicle index $n$ is dropped.
    We represent the vehicle posterior $f_{k|k}(\mathbf{s}_k) \approx \sum_{p=1}^P \mathbf{s}_{k|k}^p w_{k}^p$, where $\mathbf{s}_{k|k}^p$ is the particle, and $\sum_{p=1}^P w_{k}^p=1$ is the weight. The corrected PHD conditioned on particle sample is represented as the form of Gaussian mixture~(GM), i.e., $D_{k|k}(\mathbf{x}_k,m|\mathbf{s}_{0:k}^p) = \sum_{q=1}^{Q_k} \eta_{k|k}^{p,q}\mathcal{N}(\mathbf{x}_k;\mathbf{x}_{k|k}^{p,q}(m), \mathbf{P}_{k|k}^{p,q}(m))$, where $Q_k$ is the number of Gaussians, and  $\eta_{k|k}^{p,q}$, $\mathbf{t}_{k|k}^{p,q}$, and $\mathbf{T}_{k|k}^{p,q}$ are respectively the Gaussian weight, mean, and covariance.
    
\subsubsection{Prediction}\label{sec:Prediction}
    The vehicle particle evolves $\mathbf{s}_{k}^p =  f_\text{V}(\mathbf{s}_{k}|\mathbf{s}_{k-1}^p)$, and $w_{k|k-1}^p = w_{k-1}^p$.
    The predicted map PHD is
\begin{align}\label{eq:prevPredPHD}
    &D_{k|k-1}(\mathbf{x}_k,m|\mathbf{s}_{0:k}^p) \\
    &=  P_\text{S}(m)D_{k-1|k-1}(\mathbf{x}_{k-1},m|\mathbf{s}_{0:k-1}^p)+ D_{\text{B},k}(\mathbf{x}_{k},m|\mathbf{s}_{0:k}^p), \notag
\end{align}
    where $P_\text{S}(m)$ is the survival probability, 
    and $D_{\text{B},k}(\mathbf{x}_{k},m|\mathbf{s}_{0:k}^p)$ is the birth PHD, implemented by~\cite[Appendix~D-A]{Hyowon_TWC2020}.

\subsubsection{Correction}\label{sec:Correction}
    We correct the map PHD for $m$ as
\begin{align}
    D_{k|k}&(\mathbf{x}_k,m|\mathbf{s}_{0:k}^p) =(1-\mathsf{p}_\text{D}(\mathbf{s}_k^p,\mathbf{x}_k,m))D_{k|k-1}(\mathbf{x}_k,m|\mathbf{s}_{0:k}^p)\notag\\
    & + \sum_{\mathbf{z}\in\mathcal{Z}_k}\dfrac{\nu(\mathbf{z},\mathbf{x}'_k,m|\mathbf{s}_{0:k}^p)}{c(\mathbf{z})+\sum_{m'\in\overline{\mathcal{M}}}\int \nu(\mathbf{z},\mathbf{x}'_k,m'|\mathbf{s}_{0:k}^p) \mathrm{d}\mathbf{x}'_k},\label{eq:prevCorrPHD}
\end{align}
    where $c(\mathbf{z})$ is the clutter intensity and $\nu(\mathbf{z},\mathbf{x}_k,m|\mathbf{s}_{0:k}^p)=\mathsf{p}_\text{D}(\mathbf{s}_k^p,\mathbf{x}_k,m)D_{k|k-1}(\mathbf{x}_k,m|\mathbf{s}_{0:k}^p)g(\mathbf{z}_k|\mathbf{s}_{k}^p,\mathbf{x}_k,m)$.
    The corrected vehicle weight is $w_k^p\propto w_{k|k-1}^p \ell^p_k$, with ~\cite[Appendix~C]{Hyowon_TWC2020}
\begin{align}
    \ell^p_k
    = \sum_{\mathbf{z}\in\mathcal{Z}_k} \{   c(\mathbf{z})+\sum_{m\in\overline{\mathcal{M}}}\int \nu(\mathbf{z},\mathbf{x}_k,m|\mathbf{s}_{0:k}^p) \mathrm{d}\mathbf{x}_k\}.\label{eq:prevCorrComW}
\end{align}

\subsection{Global Map Fusion}\label{sec:MapFusion}
    Periodically, vehicle $n$ asynchronously communicates with the BS to update the global map.  
    
\subsubsection{Uplink Map Fusion at Base Station}\label{sec:uplink}
    Vehicle $n$ sends averaged map PHDs, expressed as
\begin{gather}
    \overline{D}_{k}^n(\mathbf{x}_k,m) = \sum_{p=1}^P D_{k|k}(\mathbf{x}_k,m|
    \mathbf{s}_{0:k}^{n,p}) w_k^{n,p},\label{eq:prevMFave}
\end{gather}
    along with the accumulated FOV.\footnote{The accumulated FOV is defined as $\mathcal{F}_k^n(m) = \{\mathbf{x}_k: \exists k' \in (k^{\dagger},k], \max\{p_{\text{D},k'}(\mathbf{x}_k,\mathbf{s}_{k'}^n,m)\}\geq \gamma_\text{D} \}$, where $\gamma_\text{D}$ is the threshold on target detection, close to 1, and $k^{\dagger}$ is the last communication time instant.} 
    At the BS, the fused map PHD is~\cite{Hyowon_TWC2020} 
    \begin{align}\label{eq:prevULMF}
    \widetilde{D}_k^{\text{BS}}(\mathbf{x}_{k},m) = \mathsf{F}_{\text{AA}}(\widetilde{D}_{k-1}^{\text{BS}}(\mathbf{x}_{k},m), \overline{D}_k^n(\mathbf{x}_k,m)),
\end{align}
    where $\mathsf{F}_{\text{AA}}(\cdot)$ denotes an arithmetic averaging (AA) fusion operator. Roughly speaking, given 2 PHDs $D_1$ and $D_2$, with $Q_1$ and $Q_2$ GM components indexed by $q_1$ and $q_2$, $\mathsf{F}_{\text{AA}}(D_1,D_2)$ involves the following steps: (i) computation of an $Q_1\times Q_2$ proximity matrix $\mathbf{C}$ based on the Mahalanobis distance; (ii) determining matched components $(q_1,q_2)$ between $D_1$ and $D_2$ based on the proximity matrix $\mathbf{C}$; (iii) assigning weights 
$\beta^{q_1}_1 = \beta^{q_2}_2=1/2$
to each matched pair, and assigning weights $\beta^{q_1}_1,\beta^{q_2}_2\in \{1/2,1\}$ to unmatched components\footnote{Depending on the accumulated FOV \cite{Hyowon_TWC2020}, if component is determined as false alarm, $\beta = 1/2$, otherwise, $\beta = 1$.}.
Note that missed detections are more critical than false alarms in the SLAM applications.
    To handle missed detections in the VS map, we adopt the AA approach, taking the union of the involved densities, resulting in map fusion with the minimum information loss~\cite{Gao_MOAAFusion_SPL2020}.

\subsubsection{Downlink PHD Map Update at Vehicle}\label{sec:downlink}
    Vehicle $n$ overwrites the fused map PHD as $D_{k|k}(\mathbf{x}_{k},m|\mathbf{s}_{0:k}^{n,p}) =\widetilde{D}_{k}^\text{BS}(\mathbf{x}_{k},m),~\forall p$.

\subsection{Impact of VS}
    In the presence of VS, the following effects will be observed. 
    VS on parallel trajectories to the ego vehicle lead to additional entries in the VA PHD, as they appear similar to large reflecting surfaces parallel to the ego vehicle, with slowly moving VA locations. 
    Then, SLAM accuracy is adversely affected. 

\section{Countermeasures to Track Vehicle Scatterers}\label{sec:Proposed}
\begin{figure}[t!]
\begin{centering}
	\includegraphics[width=.9\columnwidth]{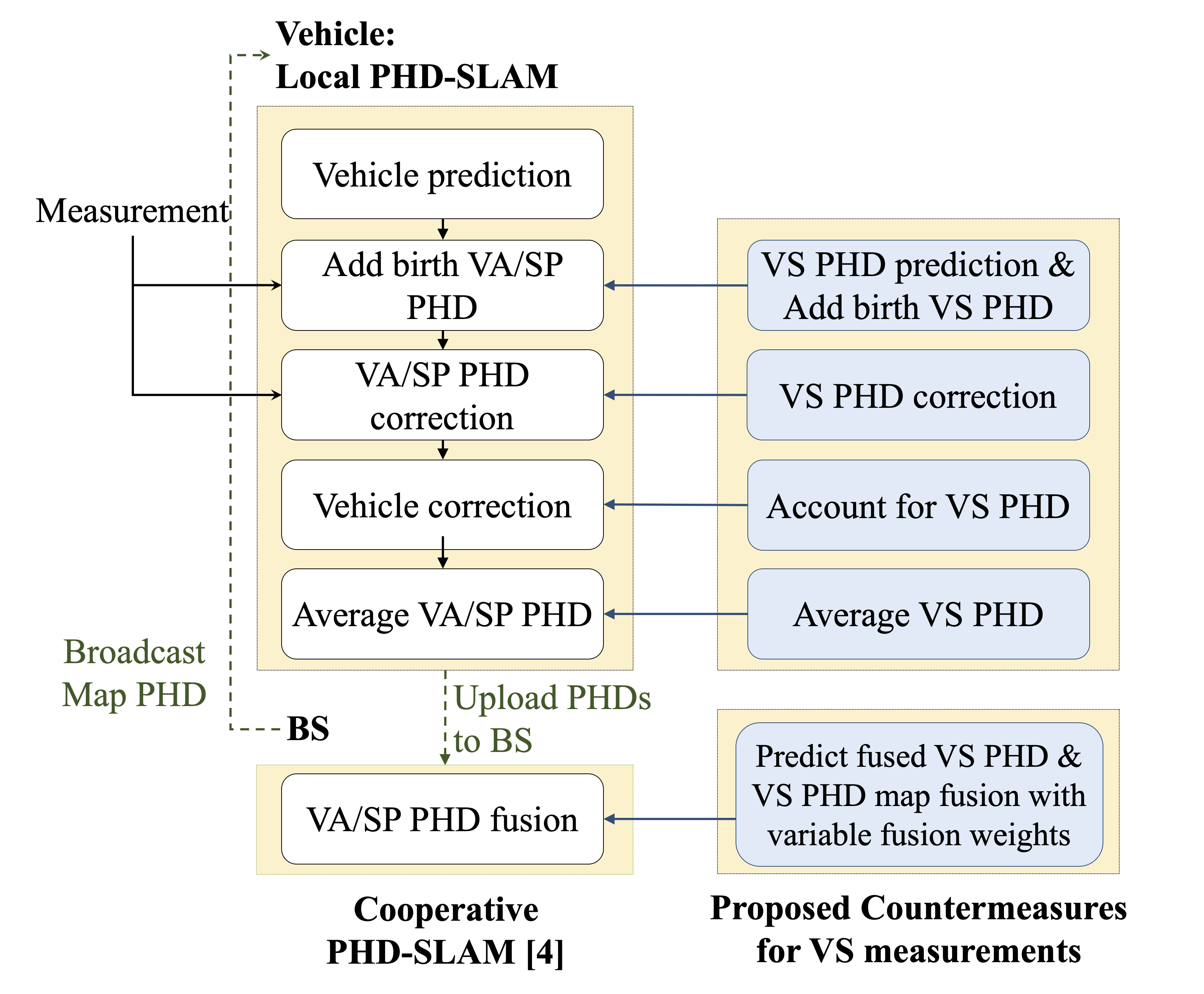}
	\caption{Graphical representation of the proposed cooperative PHD-SLAM filter.
	The directed link from countermeasures to previous PHD-SLAM indicates which countermeasure process is added on which individual process of the previous SLAM filter~\cite[Sec.~IV, V]{Hyowon_TWC2020}.}
	\label{Fig:Block}
\end{centering}
\end{figure}

    We describe the proposed countermeasure process,
    added on local PHD-SLAM (Sec.~\ref{sec:LocalPHD}) and on global map fusion (Sec.~\ref{sec:MapFusion}) as shown in Fig.~\ref{Fig:Block}. 
    We drop the target type $m$ except the process of vehicle particle correction.

\subsection{Countermeasures in Local PHD-SLAM}\label{sec:Phase1}
    To ensure a unified state definition for all targets, 
    we replace $\mathbf{x}_k$ with the target state $\mathbf{t}_k$ for the Bayesian recursion with the transition density and the measurement, and denote the corrected map PHD, conditioned on vehicle particle $\mathbf{s}_k^{n,p}$, by $D_{k|k}(\mathbf{t}_k|\mathbf{s}_{0:k}^{n,p})$, represented as the GM form.
    \subsubsection{Prediction}
    To exploit target dynamics, we include the target transition density  $f_{\text{VS}}(\mathbf{t}_{k}|\mathbf{t}_{k-1})$ in PHD map prediction:
\begin{align}\label{eq:Pro_PredPHD}
    D_{k|k-1}&(\mathbf{t}_k|\mathbf{s}_{0:k}^{n,p}) = D_{\text{B},k}(\mathbf{t}_{k}|\mathbf{s}_{0:k}^{n,p})\\ &+ P_\text{S}\int f_{\text{VS}}(\mathbf{t}_{k}|\mathbf{t}_{k-1})  D_{k-1|k-1}(\mathbf{t}_{k-1}|\mathbf{s}_{0:k-1}^{n,p})\mathrm{d}\mathbf{t}_{k-1}, \notag
\end{align}
    where $D_{\text{B},k}(\mathbf{t}_{k}|\mathbf{s}_{0:k}^{n,p})$ is the birth PHD.
    In the birth PHD, we set $\mathbf{t}_k^{n,p,q}=[(\mathbf{x}_k^{n,p,q})^\top, (\mathbf{v}_k^{n,p,q})^\top, \xi_k^{n,p,q}]^\top$ and $\mathbf{T}_k^{n,p,q}=\text{blkdiag}(\mathbf{P}_k^{n,p,q}, \mathbf{V}_k^{n,p,q}, (\sigma_\xi^{n,p,q})^2)$, where for VS, we can compute $\mathbf{x}_k^{n,p,q}$ and $\mathbf{P}_k^{n,p,q}$, similarly to SP birth PHD~\cite{Hyowon_TWC2020}, using the cubature Kalman filter (CKF).
    Due to the nonlinear measurement model, and fact that the VS velocity and turn-rate in the VS birth PHD cannot be observed, correction leads to overly concentrated covariance estimates \cite{huang2007convergence}.
    To address this, we adopt the dithering methods, mitigating approximation error for target estimates in the CKF~\cite{Gustafsson_EKF_ICCASSP2008}.
    
    
\subsubsection{Correction}
    The predicted map PHDs for all $m$ are utilized in correction, and thus the likelihood function for VS map PHD information $\int \nu(\mathbf{z},\mathbf{t}_k,\text{VS}|\mathbf{s}_{0:k}^p) \mathrm{d}\mathbf{t}_k$ is additionally considered.
    With the introduction of $\psi^p(\mathbf{z})=c(\mathbf{z})+\sum_{m\in \overline{\mathcal{M}}}\int \nu(\mathbf{z},\mathbf{x}_k,m|\mathbf{s}_{0:k}^p) \mathrm{d}\mathbf{x}_k 
    + \int \nu(\mathbf{z},\mathbf{t}_k,\text{VS}|\mathbf{s}_{0:k}^p) \mathrm{d}\mathbf{t}_k$, 
    the likelihood function is computed as $\ell^p_k=\sum_{\mathbf{z}\in\mathcal{Z}_k} \psi^p(\mathbf{z})$.
    The VS map PHD is corrected as follows:
\begin{align}
    D_{k|k}(\mathbf{t}_k,\text{VS}|\mathbf{s}_{0:k}^p) =&(1-\mathsf{p}_\text{D}(\mathbf{s}_k^p,\mathbf{x}_k,\text{VS}))D_{k|k-1}(\mathbf{t}_k,\text{VS}|\mathbf{s}_{0:k}^p)\notag\\
    & + \sum_{\mathbf{z}\in\mathcal{Z}_k}\dfrac{\nu(\mathbf{z},\mathbf{t}_k,\text{VS}|\mathbf{s}_{0:k}^p)}{\psi^p(\mathbf{z})}.\label{eq:propCorrPHD}
\end{align}

    
\subsection{Countermeasures in Global Map Fusion}\label{sec:Phase2}
    For map fusion at the BS, we first change the computation of the average map~\eqref{eq:prevMFave} by placing the accurate ego state in the averaged VS map PHD since the ego state is available with high accuracy.
    Second, we compute the fusion weights to strike an appropriate balance between missed detections and false alarms, instead of assigning the fusion weights evenly~\cite{Hyowon_TWC2020}.
    
\subsubsection{Placing the Ego Vehicle in the Average Map}
    Compared to~\eqref{eq:prevMFave}, the averaged map VS PHD at vehicle $n$ exploits the self-vehicle posterior density $f_{k|k}(\mathbf{s}_{k}^n)$.
    Then,
\begin{align}\label{eq:Pro_aveVSPHD}
    \overline{D}_{k}^n(\mathbf{t}_k) = \mathsf{F}'_{\text{AA}}( \widetilde{D}_{k}^n(\mathbf{t}_k), f_{k|k}(\mathbf{s}_{k}^n)),
\end{align}    
    where $\widetilde{D}_{k}^n(\mathbf{t}_k)=\sum_{p=1}^P D_{k|k}(\mathbf{t}_k|
    \mathbf{s}_{0:k}^{n,p},\text{VS}) w_k^{n,p}$ is the average VS PHD, similarly to~\eqref{eq:prevMFave}. 
    Here, $\mathsf{F}'_{\text{AA}}(\cdot)$ is a newly developed AA fusion operator for this global countermeasure, different from $\mathsf{F}_{\text{AA}}(\cdot)$ in Sec.~\ref{sec:MapFusion}. 
    A huge number of Gaussians are reduced by the pruning and merging~(PM) step~\cite[Table~IV]{Mullane2011}, and then
    $\widetilde{D}_{k}^n(\mathbf{t}_k)\approx \sum_{q=1}^{Q} \eta^{q}\mathcal{N}(\mathbf{t}_k;\mathbf{t}_k^{q},\mathbf{T}_k^{q})$.

    We assume that the self-vehicle posterior density $f_{k|k}(\mathbf{s}_{k}^n)$ follows the Gaussian distribution, and thus being seen as the PHD with one Gaussian with the unity weight
    $f_{k|k}(\mathbf{s}_{k}^n)\approx \mathcal{N}(\mathbf{t}_{k};{\mathbf{t}}_{k|k}^{n},{\mathbf{T}}_{k|k}^{n}).$
    To determine the fusion weights, we generate a binary proximity vector $\mathbf{c}\in \{ 0,1\}^Q$, with element $c_q=1$ if and only if
\begin{align}
    d_{\text{MSM}}(\mathcal{N}(\mathbf{x}_k;\mathbf{x}_k^{q},\mathbf{P}_k^{q}),\mathcal{N}(\mathbf{x}_{k};{\mathbf{x}}_{k|k}^{n},{\mathbf{P}}_{k|k}^{n})) < T_\text{MD}^L,
\end{align}
    otherwise, $c_q=0$. Here, $T_\text{MD}^L$ is a threshold and $d_{\text{MSM}}(\cdot)$ denotes a maximum symmetric Mahalanobis distance:
    \begin{align}
     d_{\text{MSM}}(\mathcal{N}(\mathbf{x};\bm{\mu}_1,\bm{\Sigma}_1),&\mathcal{N}(\mathbf{x};\bm{\mu}_2,\bm{\Sigma}_2))\notag \\
    & =\max(\bm{\Delta}^\top\bm{\Sigma}^{-1}_1\bm{\Delta}, \bm{\Delta}^\top\bm{\Sigma}^{-1}_2\bm{\Delta})    
    \end{align}
    with $\bm{\Delta}=\bm{\mu}_1-\bm{\mu}_2$. 
 Finally, we set the fusion weights as  
\begin{align}
    \beta_1^{q} = 
    \begin{cases}
        0, & c_q = 1,\\ 
        1, & \text{otherwise},
    \end{cases}
\end{align}
    {and $\beta_2=1$.} Here, $ c_q = 1$ indicates that self-vehicle Gaussian is matched to Gaussian $q$ in VS PHD. This ensures that matched GM components in the VS PHD are removed and replaced with a weighted vehicle posterior.
\subsubsection{Map Fusion}
    At the BS, there is no measurement, and thus the fused map PHD is predicted without the birth process
\begin{align}\label{eq:Pro_PredPHDatBS}
    \widetilde{D}_{k|k-1}^\text{BS}(\mathbf{t}_k) = P_\text{S}\int f(\mathbf{t}_{k}|\mathbf{t}_{k-1}) \widetilde{D}_{k-1}^\text{BS}(\mathbf{t}_{k-1})\mathrm{d}\mathbf{t}_{k-1}.
\end{align}
    Consequently, 
    the fused map PHD at the BS is modified as
\begin{align}\label{eq:pro_ULMF}
    \widetilde{D}_k^{\text{BS}}(\mathbf{t}_{k}) = \mathsf{F}'_{\text{AA}}( \widetilde{D}_{k|k-1}^{\text{BS}}(\mathbf{t}_{k}), \overline{D}_k^n(\mathbf{t}_k)).
\end{align}
    The PHDs in \eqref{eq:pro_ULMF} are also represented as the GM form:
\begin{align}
   \widetilde{D}_k^{\text{BS}}(\mathbf{t}_{k})  =& \sum_{q_1=1}^{Q^{\text{BS}}}  \beta_1^{q_1} \eta^{q_1}\mathcal{N}(\mathbf{t};\mathbf{t}^{q_1},\mathbf{T}^{q_1}) \\
    &+ \sum_{q_2=1}^{Q^{\text{UE}}} \beta_2^{q_2}
    \eta^{q_2} \mathcal{N}(\mathbf{t};\mathbf{t}^{q_2},\mathbf{T}^{q_2}). \notag
\end{align}
    To determine  the fusion weights $\beta_1^{q_1}$ and $\beta_2^{q_2}$, we generate a matrix $\mathbf{C}\in \{0,1\}^{Q^{\text{BS}}\times Q^{\text{UE}}}$, indicating the proximity of distance between two Gaussians, with $C_{q_1,q_2}=1$ if and only if the following two conditions are satisfied: 
\begin{align}
         d_{\text{MSM}}(\mathcal{N}(\mathbf{x};\mathbf{x}^{q_1},\mathbf{P}^{q_1}),\mathcal{N}(\mathbf{x};\mathbf{x}^{q_2},{\mathbf{P}}^{q_2})) & < T_\text{MD}^L,\\
         d_{\text{MSM}}(\mathcal{N}(\mathbf{v};\mathbf{v}^{q_1},\mathbf{V}^{q_1}),\mathcal{N}(\mathbf{v};\mathbf{v}^{q_2},{\mathbf{V}}^{q_2})) & < T_\text{MD}^V,
\end{align}
    where $T_\text{MD}^V$ is a threshold\footnote{The MSM metric can also be used in a Hungarian algorithm.}.  
    The fusion weights are finally assigned as follows:
\begin{itemize}
    \item \textit{Matched Targets}~(i.e., ${C}_{q_1,q_2}=1$): The condition notes that both two targets with states $\mathbf{t}^{q_1}$ and $\mathbf{t}^{q_2}$ are matched, and then two VS densities need to be merged.
    To weigh the matched densities according to their covariance,
    the Gaussian uncertainty is determined by $\rho = \text{trace}(\mathbf{T})/\text{dim}(\mathbf{t})$~\cite{DN_statistical}.
    Then, we compute the weights as
\begin{align}\label{eq:CompAFW}
    \beta_1^{q_1} = \dfrac{{1}/{\rho^{q_1}}}{{1}/{\rho^{q_1}} + {1}/{\rho^{q_2}}}, &&
    \beta_2^{q_2} = \dfrac{{1}/{\rho^{q_2}}}{{1}/{\rho^{q_1}} + {1}/{\rho^{q_2}}}.
\end{align}
    \item \textit{Unmatched Targets}~(i.e., ${C}_{q_1,q_2}=0$): The condition notes that the targets $\mathbf{t}^{q_1}$ and $\mathbf{t}^{q_2}$ are unmatched.
    If $\sum_{q_1}C_{q_1,q_2} = 0$,
    then the Gaussian $q_2$ is not be associated with any Gaussian $q_1$, 
    indicating that Gaussian $q_2$ could be a newly detected target or false alarm.
    To avoid the risk of missed detection, we set $\beta_2^{q_2}=1$.
    If $\sum_{q_2}C_{q_1,q_2} = 0$, then Gaussian $q_1$ is not associated with any Gaussian $q_2$, indicating that Gaussian $q_1$ could be the previously detected target or previous false alarm, possibly.
    Here, we can make a decision based on the accumulated FOV.
    If $\mathbf{t}^{q_1} \notin \mathcal{F}_k^n(m)$, then Gaussian $q_1$ is decided as the previously detected target by other vehicles and thus we set $\beta_1^{q_1} = 1$ for keeping Gaussian $q_1$ as the true target.
    If $\mathbf{t}^{q_1} \in \mathcal{F}_k^n(m)$, then Gaussian $q_1$ is decided as a previous false alarm. We set $0< \beta_1^{q_1} <1$  (e.g., $0.25$) for reducing the weight of the Gaussian $q_1$ density generated by possible previous false alarm, while simultaneously preventing missed detections. 
\end{itemize}

\section{Simulation Results and Discussions}\label{sec:Results}
\subsection{Simulation Setup}\label{sec:SimulSetup}
    To  demonstrate the developed PHD-SLAM filter with proposed countermeasures to handle VSs, we consider a scenario that VS measurements are added into mmWave radio propagation environment~\cite{Hyowon_TWC2020}.
    We consider two vehicles which move parallel on the circular road.
    The same values in~\cite[Sec.~VI-A]{Hyowon_TWC2020} are adopted for the parameters.\footnote{Process noise $\mathbf{q}_{k}$; time interval $\Delta$, measurement noise covariance $\mathbf{R}$; BS location $\mathbf{x}_\text{BS}$, VA locations $\mathbf{x}_\text{VA,1}$, $\mathbf{x}_\text{VA,2}$, $\mathbf{x}_\text{VA,3}$, $\mathbf{x}_\text{VA,4}$; threshold for PM steps (for the reduction of Gaussian components and the weighted sum of Gaussians); threshold for target detection $T_\text{VA}$, $T_\text{VS} = T_\text{SP}$;
    birth weight; Poisson mean for clutter measurement; landmark visibility and FOV range $r_\text{SP}=r_\text{VS}=50~\text{m}$.}
    
    For dynamics of vehicle states~\eqref{eq:VehicleDynamic} and VSs~\eqref{eq:TragetDynamic}, we respectively adopt~\cite[eq.~(25)]{Hyowon_TWC2020} in polar coordinates and~\cite[Sec.~V-B]{Li_Dynamics_TAES2003} in Cartesian coordinates, rendering the VS state identifiable over time. 
    We set $\bar{\mathbf{q}}_k=[1, 1, 0.1, 3, 3, 0.1, 0.05]^\top$, with units m, m, m, m/s, m/s, m/s, rad/s.
    For generating $\mathbf{v}^i$ and $\xi^i$ in $D_{\text{B},k}(\mathbf{t}_k|\mathbf{s}_{0:k}^{n,p})$ of~\eqref{eq:Pro_PredPHD}, we set $\mathbf{V}=\text{diag}([100, 100, 0.09])$ and $\sigma_\xi = \pi/2$, and 
    $\mathbf{v}^i \sim \mathcal{N}(\bm{0},\mathbf{V})$ and $\xi^i \sim \mathcal{U}(0,2\pi]$.
    In VS prediction,  $\mathbf{T}_d=\text{diag}([9,9,0.09,5,5,0.09,0.18])$ is added into each covariance of VS Gaussians during~\eqref{eq:Pro_PredPHD}.
    The vehicle states are initialized as $\mathbf{s}^{1}_0 = [70.73, 0, 0, \pi/2, 22.22, \pi/10, 300]^\top$ and ${\mathbf{s}^{2}_0 = [60.73, 0, 0,\pi/2, 19.08, \pi/10, 300]^\top}$, with units m, m, m, rad, m/s, rad/s, and m.
    The initial prior of the vehicle state is sampled from $\mathcal{N}(\mathbf{s}_0;\mathbf{s}_0^n,\mathbf{S}_0^n)$, where $\mathbf{S}_0^n=\text{diag}(0.3^2, 0.3^2, 0, 0.1^4, 0, 0, 0.3^2)$.
    The longitudinal velocity $\zeta_k^n$, rotational velocity $\xi_k^n$ are assumed to be known since the knowledge of inertial sensor of board is available.
    Four SPs are located at $[\pm 55, \pm 55, z_\text{SP}]^\text{T}$ m, where $z_\text{SP}\sim \mathcal{U}(0,40)$.
    We set $p_{\text{D}}=0.95$ within the FOV and $P(m) = 0.99$.
    
    After the PHD map fusion step~(Sec.~\ref{sec:uplink}), the PM step in~\cite[Table II]{Vo2006} is used, and the PM step in~\cite[Table~IV]{Mullane2011} is used in averaging PHD map~\eqref{eq:prevMFave}.
    In the PM steps of averaging PHD map and PHD map fusion, pruning threshold is set to 0.1, preventing the false Gaussian with large covariance from diluting the true Gaussian with small covariance.
    Both thresholds $M_\text{MD}^L$ and $M_\text{MD}^V$ are set to 20.
    In asynchronous map fusion, vehicles communicate with the BS every 2 time steps, with vehicle 1 and vehicle 2 respectively starting at time 5 and 6.
    For representing each vehicle state, $P=300$ particles are used, and results are averaged over $10$ Monte Carlo runs. To evaluate the mapping accuracy, the average of the generalized optimal subpattern assignment (GOSPA)~\cite{RahmathullahGFS:2017} distance is utilized.

\begin{figure}[t!]
\begin{centering}
	\subfloat[\label{Fig:GOSPA_VS}]
	{\includegraphics[width=.9\columnwidth]{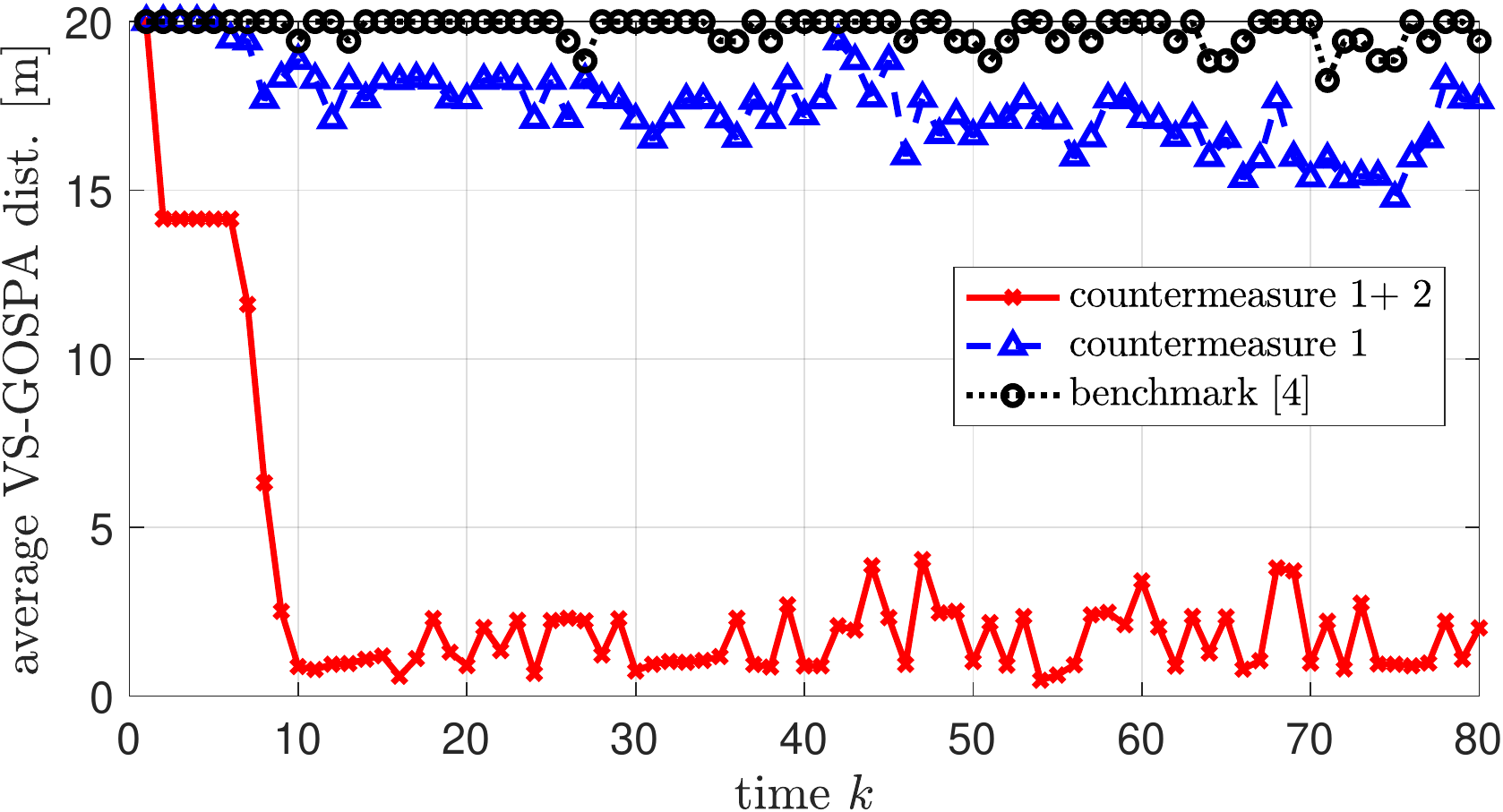}}\hfill
	\subfloat[\label{Fig:GOSPA_VA}]
	{\includegraphics[width=.9\columnwidth]{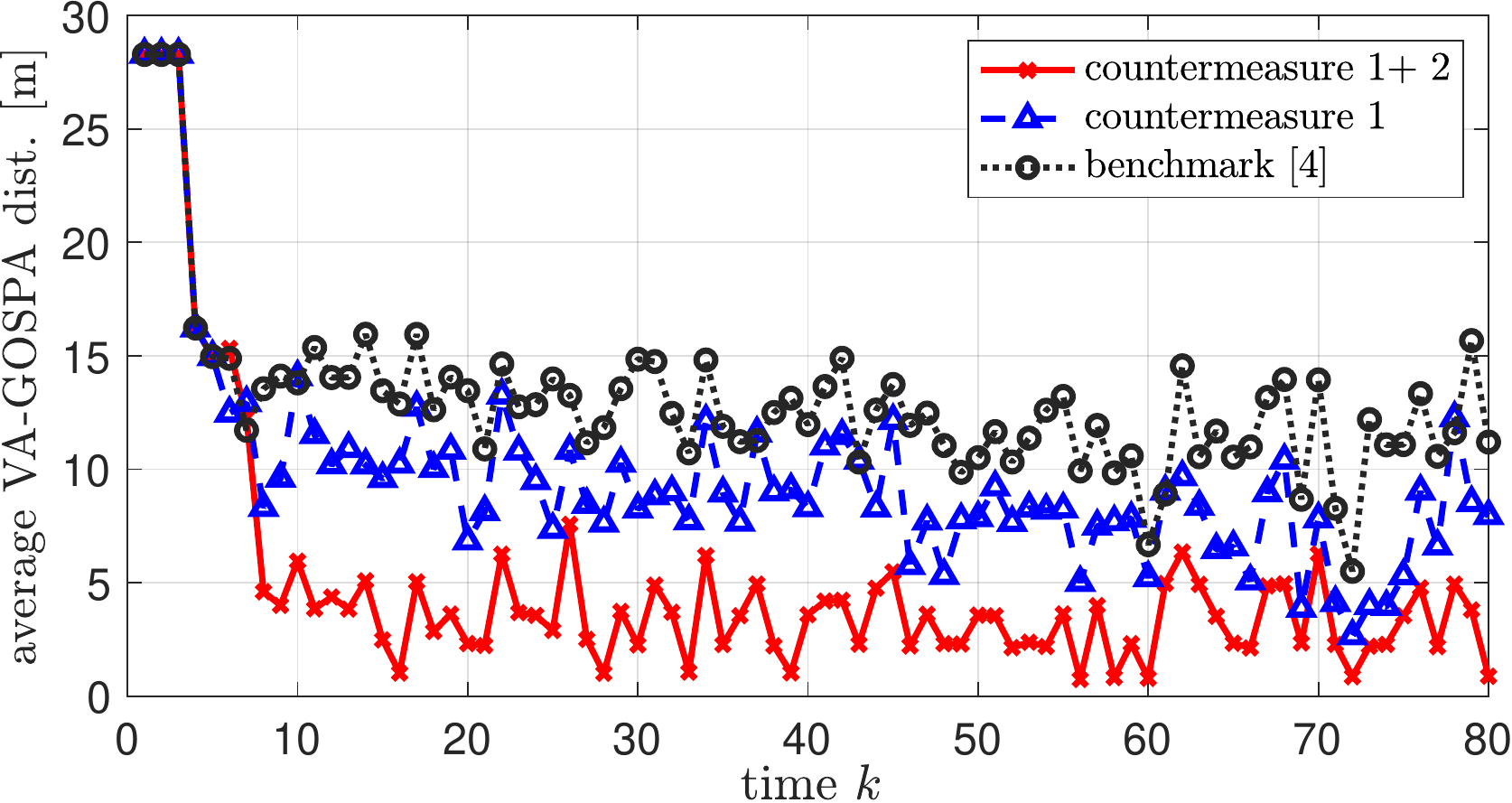}}\hfill
	\caption{Mapping performance: Average GOSPA of (a) VS and (b) VA.}
	\label{Fig:GOSPA}
	\par
\end{centering}
\end{figure}

\begin{figure}[t!]
\begin{centering}
	{\includegraphics[width=.9\columnwidth]{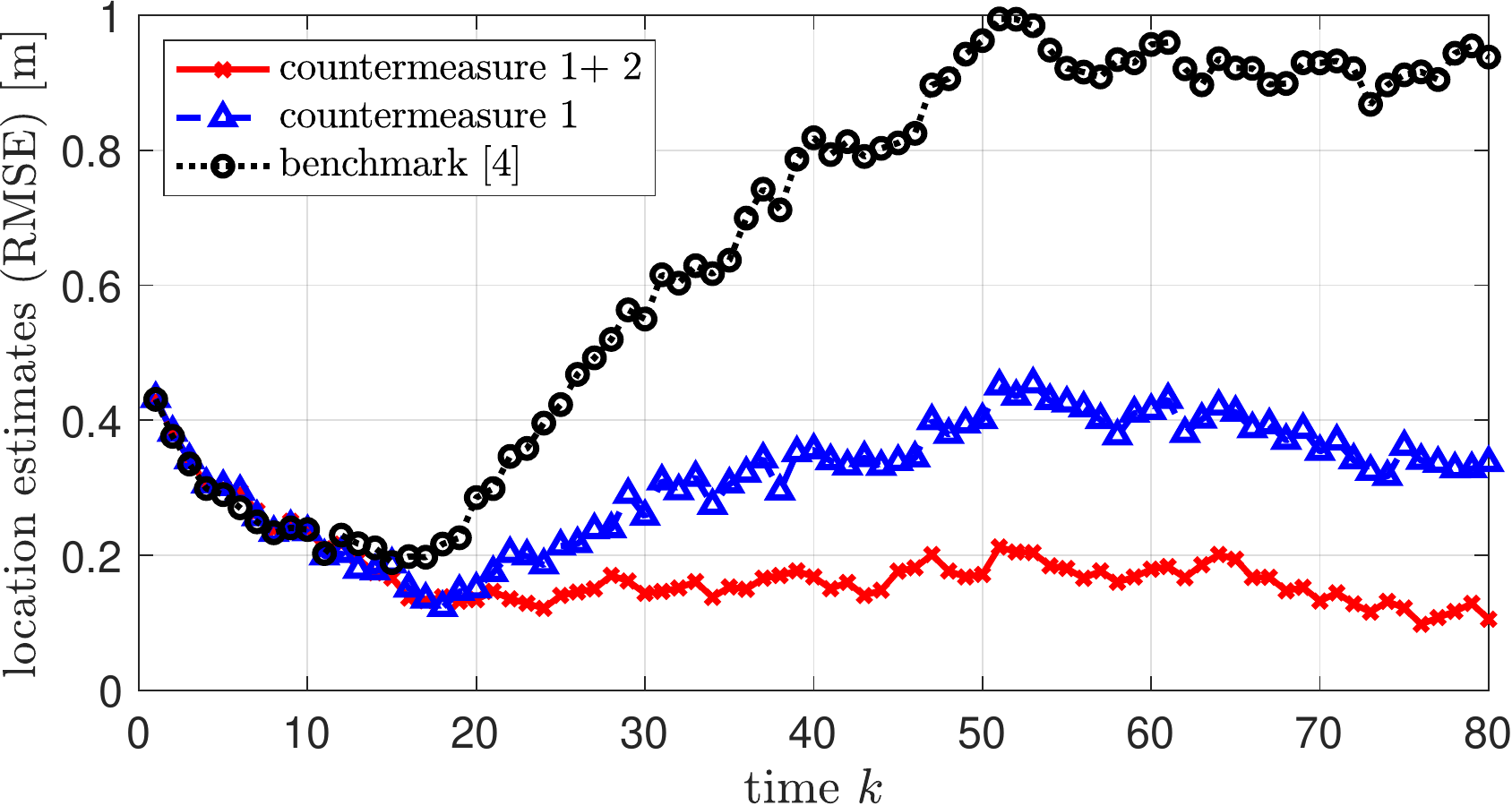}}
	\caption{Localization performance: Average error of vehicle state estimates in terms of location RMSE.}
	\label{Fig:VehicleStateEst}
\end{centering}
\end{figure}


\subsection{Results and Discussion}
\subsubsection{Mapping} 
    Fig.~\ref{Fig:GOSPA} shows the mapping accuracy of the proposed filter, evaluated by averaging GOSPA distances.
    The VS measurement cannot be handled by the MM-PHD-SLAM filter from~\cite{Hyowon_TWC2020}, and thus false detection appears in the VA map PHD as shown in Fig.~\ref{Fig:GOSPA_VA} and inevitably the VS target is missed as shown in Fig.~\ref{Fig:GOSPA_VS}.
    The developed MM-PHD-SLAM filter with proposed countermeasure 1 (see, Sec.~\ref{sec:Phase1}) slightly improves mapping accuracy, and the VS measurement is still not handled since the VS velocity cannot be obtained without the self-vehicle posterior density.
    However, we confirm that the proposed MM-PHD-SLAM filter (i.e., with both countermeasures) handles the challenge (false alarms in the VA map PHD and missed detections in the VS map PHD, due to VS measurements) by virtue of the proposed fused map PHD.
    
\subsubsection{Localization} 
    Fig.~\ref{Fig:VehicleStateEst} shows the accuracy of the vehicle state estimates, evaluated by root mean square error~(RMSE) for the location (similar behavior was observed for the clock bias and heading).
    For the filter from~\cite{Hyowon_TWC2020} and the proposed MM-PHD-SLAM filter, we observe similar trends as in mapping,
    due to the fact that vehicle and map are correlated at every time step.
    The RMSEs are similar at first. However, after starting map fusion at time 5, and after a few time steps, the fused map is becoming informative, then the RMSEs diverge.
    

\section{Conclusions}\label{sec:Conclusions}
    We have proposed countermeasures to handle moving VSs, in cooperative mmWave radio PHD-SLAM.
    We demonstrate that without these countermeasures, standard PHD-SLAM fails, due to false targets raised by the VS. 
    We confirmed that the proposed filter can track both moving VSs and static landmarks, while simultaneously localize the vehicles.
    

\bibliographystyle{IEEEtran}
\bibliography{bibliography}

\begin{thebibliography}{10}
\providecommand{\url}[1]{#1}
\csname url@samestyle\endcsname
\providecommand{\newblock}{\relax}
\providecommand{\bibinfo}[2]{#2}
\providecommand{\BIBentrySTDinterwordspacing}{\spaceskip=0pt\relax}
\providecommand{\BIBentryALTinterwordstretchfactor}{4}
\providecommand{\BIBentryALTinterwordspacing}{\spaceskip=\fontdimen2\font plus
\BIBentryALTinterwordstretchfactor\fontdimen3\font minus
  \fontdimen4\font\relax}
\providecommand{\BIBforeignlanguage}[2]{{%
\expandafter\ifx\csname l@#1\endcsname\relax
\typeout{** WARNING: IEEEtran.bst: No hyphenation pattern has been}%
\typeout{** loaded for the language `#1'. Using the pattern for}%
\typeout{** the default language instead.}%
\else
\language=\csname l@#1\endcsname
\fi
#2}}
\providecommand{\BIBdecl}{\relax}
\BIBdecl

\bibitem{Henk_5GmmWavePosi_WC2018}
H.~Wymeersch, G.~Seco-Granados, G.~Destino, D.~Dardari, and F.~Tufvesson,
  ``{5G} mm-{W}ave positioning for vehicular networks,'' \emph{IEEE Wireless
  Commun.}, vol.~24, no.~6, pp. 80--86, Dec. 2018.

\bibitem{Erik_BPSLAM_TWC2019}
E.~Leitinger, F.~Meyer, F.~Hlawatsch, K.~Witrisal, F.~Tufvesson, and M.~Z. Win,
  ``A belief propagation algorithm for multipath-based {SLAM},'' \emph{IEEE
  Trans. Wireless Commun.}, vol.~18, no.~12, pp. 5613--5629, Dec. 2019.

\bibitem{Rico_BPSLAM_JSTSP2019}
R.~Mendrzik, F.~Meyer, G.~Bauch, and M.~Z. Win, ``{Enabling situational
  awareness in millimeter wave massive MIMO systems},'' \emph{IEEE J. Sel.
  Topics Signal Process.}, vol.~13, no.~5, pp. 1196--1211, Sep. 2019.

\bibitem{Hyowon_TWC2020}
H.~Kim, K.~Granstr\"{o}m, L.~Gao, G.~Battistelli, S.~Kim, and H.~Wymeersch,
  ``{5G mmWave cooperative positioning and mapping using multi-model PHD},''
  \emph{IEEE Trans. Wireless Commun.}, vol.~19, no.~6, pp. 3782--3795, Mar.
  2020.

\bibitem{Pasha_MMPHD_TAES2009}
S.~Pasha, B.-N. Vo, H.~Tuan, and W.-K. Ma, ``A {G}aussian mixture {PHD} filter
  for jump {M}arkov system models,'' \emph{IEEE Trans. Aerosp. Electron.
  Syst.}, vol.~45, no.~3, pp. 919--936, Jul. 2009.

\bibitem{Karl_MMPHD_FUSION2014}
K.~Gran\-str{\"o}m, S.~Reuter, D.~Meissner, and A.~Scheel, ``{A multiple model
  PHD approach to tracking of cars under an assumed rectangular shape},'' in
  \emph{Proc. 17th Int. Conf. Inf. Fusion (FUSION)}, Salamanca, Spain, Jul.
  2014.

\bibitem{mahler_book_2007}
R.~Mahler, \emph{Statistical Multisource-Multitarget Information Fusion}.\hskip
  1em plus 0.5em minus 0.4em\relax Norwood, MA, USA: Artech House, 2007.

\bibitem{bas2017dynamic}
C.~U. Bas, R.~Wang, S.~Sangodoyin, S.~Hur, K.~Whang, J.~Park, J.~Zhang, and
  A.~F. Molisch, ``{Dynamic double directional propagation channel measurements
  at 28 GHz},'' \emph{arXiv preprint arXiv:1711.00169}, 2017.

\bibitem{Henk_FOV_6G2020}
H.~Wymeersch and G.~Seco-Granados, ``Adaptive detection probability for
  mm{W}ave 5{G} {SLAM},'' in \emph{6G Wireless Summit (6G SUMMIT)}, 2020.

\bibitem{Yu_DiffusePMBM_Sensors2020}
Y.~Ge, H.~Kim, F.~Wen, L.~Svensson, S.~Kim, and H.~Wymeersch, ``Exploiting
  diffuse multipath in {5G SLAM},'' \emph{arXiv preprint arXiv:2006.15603},
  2020.

\bibitem{Gao_MOAAFusion_SPL2020}
L.~Gao, G.~Battistelli, and L.~Chisci, ``Multiobject fusion with minimum
  information loss,'' \emph{IEEE Signal Process. Lett.}, vol.~27, pp. 201--205,
  Jan. 2020.

\bibitem{huang2007convergence}
S.~Huang and G.~Dissanayake, ``{Convergence and consistency analysis for
  extended Kalman filter based SLAM},'' \emph{IEEE Trans. Robot.}, vol.~23,
  no.~5, pp. 1036--1049, Oct. 2007.

\bibitem{Gustafsson_EKF_ICCASSP2008}
F.~Gustafsson and G.~Hendeby, ``On nonlinear transformations of stochastic
  variables and its application to nonlinear filtering,'' in \emph{in Proc.
  IEEE Int. Conf. Acoust. Speech Signal Process. (ICASSP’08)}.\hskip 1em plus
  0.5em minus 0.4em\relax IEEE, 2008, pp. 3617--3620.

\bibitem{Mullane2011}
J.~Mullane, B.-N. Vo, M.~D. Adams, and B.-T. Vo, ``{A random-finite-set
  approach to Bayesian SLAM},'' \emph{IEEE Trans. Robot.}, vol.~27, no.~2, pp.
  268--282, Apr. 2011.

\bibitem{DN_statistical}
D.~Paindaveine, ``A canonical definition of shape,'' \emph{Statist. \& Probab.
  Lett.}, vol.~78, no.~14, pp. 2240--2247, Feb. 2008.

\bibitem{Li_Dynamics_TAES2003}
X.~Rong-Li and V.~Jilkov, ``Survey of maneuvering target tracking: {P}art {I}.
  {D}ynamic models,'' \emph{IEEE Trans. Aerosp. Electron. Syst.}, vol.~39,
  no.~4, pp. 1333--1364, Oct. 2003.

\bibitem{Vo2006}
B.-N. Vo and W.-K. Ma, ``The {G}aussian mixture probability hypothesis density
  filter,'' \emph{IEEE Trans. Signal Process.}, vol.~54, no.~11, pp.
  4091--4104, Nov. 2006.

\bibitem{RahmathullahGFS:2017}
A.~S. Rahmathullah, A.~F. Garc\'{i}a~Fern\'{a}ndez, and L.~Svensson,
  ``Generalized optimal sub-pattern assignment metric,'' in \emph{Proc. 20th
  Int. Conf. Inf. Fusion (FUSION)}, Xian, China, Jul. 2017, pp. 1--8.

\end{thebibliography}
\end{document}